\documentclass[aps,prd,superscriptaddress,twocolumn,showpacs,floatfix,10pt]{revtex4-1}
\usepackage{color}
\usepackage{ulem}
\usepackage{enumerate}
\usepackage{amsmath}
\usepackage{graphicx}

\def \be {\begin{equation}}
\def \ee {\end{equation}}
\def \ba {\begin{array}}
\def \ea {\end{array}}
\def \bea {\begin{eqnarray}}
\def \eea {\end{eqnarray}}

\def \ble {\begin{widetext}\begin{equation}}
\def \ele {\end{equation}\end{widetext}}
\def \blea {\begin{widetext}\begin{eqnarray}}
\def \elea {\end{eqnarray}\end{widetext}}

\def \nn {\nonumber}
\def \ketpsi {|\psi\rangle}
\def \brapsi {\langle \psi |}
\def \ketphi {|\phi\rangle}
\def  \braphi {\langle \phi|}
\def \trans {\mathcal{T}^{\psi|\phi}}
\def \transA {\mathcal{T}_A^{\psi|\phi}}
\def \transAc {\mathcal{T}_A^{\phi|\psi}}
\def \blea {\begin{widetext}\begin{eqnarray}}
\def \elea {\end{eqnarray}\end{widetext}}
\def \mO {\mathcal{O}}
\def \mtO {\tilde{\mathcal{O}}}
\def \mA {\mathcal{A}}

\usepackage[colorlinks=true,linkcolor=red,citecolor=blue,urlcolor=blue,bookmarks]{hyperref}
\usepackage{eufrak}

\def \be {\begin{equation}}
\def \ee {\end{equation}}
\def \ba {\begin{array}}
\def \ea {\end{array}}
\def \bea {\begin{eqnarray}}
\def \eea {\end{eqnarray}}
\def \nn {\nonumber}

\def \a {\alpha}
\def \b {\beta}

\def \l {\lambda}

\def \r {\rho}

\def \cH {\mathcal H}

\def \cS {\mathcal S}
\def \cT {\mathcal T}

\def \p {\partial}

\def \f {\frac}

\def \sr {\sqrt}
\def \td {\tilde}

\def \lag {\langle}
\def \rag {\rangle}

\def \ep {\mathrm{e}}
\def \ii {\mathrm{i}}

\def \tr {\textrm{tr}}

\def \and {{~\textrm{and}~}}

\def \ttrans {{\tilde{\mathcal{T}}^{\psi|\phi}}}

\begin{document}

\title{Sum rule for the pseudo-R\'enyi entropy}

\author{Wu-zhong Guo}
\email{wuzhong@hust.edu.cn}
\affiliation{School of Physics, Huazhong University of Science and Technology,
Luoyu Road 1037, Wuhan, Hubei 430074, China}

\author{Jiaju Zhang}
\email{jiajuzhang@tju.edu.cn}
\affiliation{Center for Joint Quantum Studies and Department of Physics, School of Science,
Tianjin University, 135 Yaguan Road, Tianjin 300350, China}

\begin{abstract}

  By generalizing the density matrix to a transition matrix between two states, represented as $|\phi\rangle$ and $|\psi\rangle$, one can define the pseudoentropy analogous to the entanglement entropy. In this paper, we establish an operator sum rule that pertains to the reduced transition matrix and reduced density matrices corresponding to the superposition states of $|\phi\rangle$ and $|\psi\rangle$. It is demonstrated that the off-diagonal elements of operators can be correlated with the expectation value in the superposition state. Furthermore, we illustrate the connection between the pseudo-R\'enyi entropy and the R\'enyi entropy of the superposition states. We provide proof of the operator sum rule and verify its validity in both finite-dimensional systems and quantum field theory.  We additionally demonstrate the significance of these sum rules in gaining insights into the physical implications of transition matrices, pseudoentropy, and their gravity dual.

\end{abstract}

\maketitle

\section{Introduction}

In many research fields, we encounter the handling of off-diagonal elements in certain states. To understand the quantum decoherence the off-diagonal elements of observables in the pointer states play an important role \cite{Zurek:1981xq,Zurek:1982ii}. In the eigenstate thermalization hypothesis \cite{Deutsch:1991msp,Srednicki:1994mfb} one concerns about the matrix elements of local observables between energy eigenstates. In the field of quantum weak value the key quantity is the  expectation value of operator between the so-called preselection and postselection states \cite{Aharonov:1988xu,Dressel:2014hlj}.

To express the off-diagonal matrix elements between two states, say $|\phi\rangle$ and $|\psi\rangle$, one can define the unnormalized transition matrix as $\tilde{\mathcal{T}}^{\psi|\phi} := \ketpsi \braphi$. The off-diagonal elements of local observables $\mathcal{A}$ can be expressed as
\bea
\langle \phi|\mathcal{A} |\psi\rangle = \tr (\ttrans \mathcal{A}).
\eea
Here, the transition matrix $\ttrans$ plays a role similar to that of the density matrix. In fact, in the special case where $|\phi\rangle=|\psi\rangle$, it becomes the density matrix. Thus, $\ttrans$ can be considered a generalization of the density matrix. Many quantities can be rephrased in terms of the transition matrices. For instance, the expectation value of an operator $\mathcal{O}$ in the state $\rho=|\phi\rangle \langle \phi|$ is given by $\tr(\rho \mathcal{O})$. It can also be understood as $\tr(\ttrans)$ with $|\psi\rangle:=\mathcal{O}|\phi\rangle$.

If a system can be partitioned into a subsystem $A$ and its complementary counterpart $\bar A$, similar to the reduced density matrix, one can establish the reduced transition matrix of $A$ as $\transA=\tr_{\bar A} \trans$, where we have defined the normalized transition matrix
\bea\label{transitionmatrix}
\trans := \frac{\ketpsi \braphi}{\braphi \psi\rangle},
\eea
assuming $|\phi\rangle$ and $|\psi\rangle$ are non-orthogonal. The above idea is explored in quantum field theory in a recent paper \cite{Nakata:2020luh}, see also \cite{Murciano:2021dga}. By definition, one can see that $\transA$ is generally a non-Hermitian matrix.
It was shown that the reduced transition matrices contain additional theoretical information and lead to intriguing results.

In this paper, we present a sum rule that connects the reduced transition matrix $\cT_A^{\psi|\phi}$ and the reduced density matrix of the superposition state composed of $|\psi\rangle$ and $|\phi\rangle$. To accomplish this, we introduce a series of states denoted by $|\xi(c)\rangle$, defined as
\bea\label{superposition}
|\xi(c)\rangle=\mathcal{N}(c)(\ketphi+c \ketpsi),
\eea
where $\mathcal{N}(c)=1/\sqrt{1+|c|^2+c \langle \phi\ketpsi+c^* \langle \psi \ketphi}$ ensures the normalization of $\lag \xi(c) | \xi(c) \rag = 1$.
We use $\rho_A(c):=\tr_{\bar A} |\xi(c)\rangle \langle \xi(c)|$ to represent the reduced density matrix.
The key finding of this paper is the sum rule that establishes a connection between $(\transA)^n$ and $[ \rho_A(c)]^n$ as
\bea\label{operatorsumrule}
 (\transA)^n=\sum_{c\in \mathcal{S}} a(c) [ \rho_A(c) ]^n ,
\eea
where $a(c)$ are some specific $n$-dependent constants, and the index $c$ belongs to a given set denoted by $\mathcal{S}$.

In the next section, we will demonstrate that one of the choices is $\mathcal{S}=\big\{\ep^{\frac{2\pi \ii}{2n+1}k} \big| k=0,1,2,\cdots,2n\big\}$. For this particular set, we define
\bea\label{ak}
&&a_k:= a(\ep^{\frac{2\pi \ii}{2n+1}k}) =\frac{\ep^{-\frac{2\pi \ii}{2n+1}kn}}{(2n+1) {{\mathcal{N}}_k^{n}}},\\
&&{\mathcal{N}}_k:=\mathcal{N}(\ep^{\frac{2\pi\ii}{2n+1}k})^2 \langle\phi |\psi \rangle.
\eea
As a result, the operator sum rule takes the form
\bea\label{opsummain}
(\transA)^n=\sum_{k=0}^{2n} a_k [\rho_A(k)]^n,
\eea
where $\rho_A(k):= \rho_A(\ep^{\frac{2\pi \ii}{2n+1}k})$.

Using the operator sum rule (\ref{opsummain}), we can establish a sum rule for off-diagonal matrix elements. The off-diagonal matrix elements of a set of operators $\{\mathcal{A}_j\}$ $(j=1,...,m)$  in subsystem $A$ would have the following interesting relations. If $|\phi\rangle$ and $|\psi\rangle$ are non-orthogonal, for $m\le n$ we have
\bea\label{correlator_sumrule_nonorth}
\prod_{j=1}^{m}\frac{\langle \phi| \mA_j|\psi\rangle}{\langle \phi| \psi\rangle}=\sum_{k=0}^{2n}a_k \prod_{j=1}^{m}\langle \xi_k| \mA_j|\xi_k\rangle,
\eea
where $|\xi_k\rangle:= |\xi(\ep^{\frac{2\pi \ii}{2n+1}k})\rangle$.
If $|\phi\rangle$ and $|\psi\rangle$ are orthogonal, we have
\bea\label{correlator_sumrule_orth}
\prod_{j=1}^{n}\langle \phi| \mA_j|\psi\rangle=\sum_{k=0}^{2n}a'_k \prod_{j=1}^{n}\langle \xi_k| \mA_j|\xi_k\rangle,
\eea
where $a_k'=\frac{2^n}{2n+1}\ep^{-\frac{2\pi \ii}{2n+1}kn}$.

Based on the similarity between the transition matrix and the density matrix, the authors in \cite{Nakata:2020luh} introduce a concept called pseudoentropy. For the reduced transition matrix $\transA$ the definition is
\bea
S(\transA)= - \tr_A ( \transA \log \transA ).
\eea
To evaluate the pseudoentropy, the concept of pseudo-R\'enyi entropy is introduced
\bea\label{pseudoRenyi}
S^{(n)}(\transA)= \frac{\log \tr_A (\transA)^n}{1-n},
\eea
where $n$ is an integer greater than or equal to 2. By analytically continuing $n$, the pseudoentropy can be expressed as $S(\transA)=\lim\limits_{n\to 1} S^{(n)}(\transA)$.
In practice, the replica trick is commonly used to evaluate pseudo-R\'enyi entropy in QFTs. Although the calculation method for pseudo-R\'enyi entropy is similar to that of R\'enyi entropy, there are relatively few analytical results available for pseudo-R\'enyi entropy. This scarcity is attributed to the transition matrix involving two distinct states and being typically non-Hermitian, which makes the calculations more intricate.

After taking the trace on both sides of equation (\ref{opsummain}), we obtain an interesting relation between the R\'enyi entropy and pseudo-R\'enyi entropy
\bea\label{sumrulegeneral}
\ep^{(1-n)S^{(n)}(\transA)}=\sum_{k=0}^{2n} a_k \ep^{(1-n)S^{(n)}(\rho_A(k))}.
\eea
The discrete Fourier transform of a sequence of $N$ complex numbers $\{ x_k\} $ with $k\in \{0,...,N-1\}$ could be denoted by $F[x_k](s)=\sum_{k=0}^{N-1}\ep^{-\frac{2\pi \ii}{2n+1}ks}x_k$.
The results (\ref{correlator_sumrule_nonorth}), (\ref{correlator_sumrule_orth}) and (\ref{sumrulegeneral}) show the off-diagonal elements and the pseudo-R\'enyi entropy can be associated with the discrete Fourier transform with $N=2n+1$ and $s=n$.
Therefore, the calculation of these quantities can be transformed into evaluating diagonal elements and the R\'enyi entropy for the superposition states and performing a discrete Fourier transformation.
It is worth noting that the above results are trivially satisfied for the special case where $\ketpsi=\ketphi$.

The subsequent sections of the paper are structured in the following manner.
Section~\ref{sectionSig} explores the importance of the sum rule.
In section~\ref{sectionPro}, we give a proof for the sum rule.
We discuss the sum rule in quantum field theory (QFT) and conformal field theory (CFT) in sections~\ref{sectionQFT} and \ref{sectionCFT}, respectively.
We present examples of the sum rule in section~\ref{sectionSim}.
Moreover, section~\ref{sectionApp} highlights several applications of the sum rule.
In section~\ref{sectionTen}, we study a tentative sum rule of the pseudoentropy.
Section~\ref{sectionAlt} exhibits alternate forms of the sum rule.
Analogous to the sum rule, section~\ref{sectionGen} demonstrates the generalized R\'enyi entropy expressed as derivatives of R\'enyi entropy.
We conclude with discussion in section~\ref{sectionDis}.

\section{The significance of the sum rules} \label{sectionSig}

Before delving into the details of proving the sum rule, let us first explore its significance.
It is well known that any non-Hermitian operator can be expressed as a linear combination of two Hermitian operators. Our sum rule (\ref{opsummain}) illustrates a distinct and generalized version of a similar statement. The sum rules involving the off-diagonal elements (\ref{correlator_sumrule_nonorth}), (\ref{correlator_sumrule_orth}) and pseudo-R\'enyi entropy (\ref{sumrulegeneral}) represent two different aspects of the operator relation.

As mentioned in the introduction, in many situations one needs to deal with the off-diagonal elements in some Hilbert space, denoted by the $\cH$ with the orthonormal basis $\mathcal{B}:=\{ |\mathfrak{m}_i\rangle \}$ ($i=0,1,...$). Consider the off-diagonal elements $\langle \mathfrak{m}_0| \mA_j |\mathfrak{m}_i\rangle$ ($i\ge 1$) for operator $\mathcal{A}$. One can construct a new set of nonothogonal basis $\mathcal{B}^{k}:= \{|\mathfrak{m}'_i(k)\rangle\}$ with $|\mathfrak{m}'_i(k)\rangle:=\mathcal{N}(k) (|\mathfrak{m}_0\rangle +\ep^{\frac{2\pi\ii}{2n+1}k}|\mathfrak{m}_i\rangle)$ ($i=0,1,...$), where $\mathcal{N}(k)$ is the normalization constant. Note that $\mathcal{B}^k$ is constructed by a rotation of the basis $|\mathfrak{m}_i\rangle$ ($i\ge 1$) with respect to $|\mathfrak{m}_0\rangle$.
Eq. (\ref{correlator_sumrule_orth}) suggests the product of the off-diagonal elements  $\prod_{j=1}^n\langle \mathfrak{m}_0| \mA_j |\mathfrak{m}_i\rangle$ in the basis $\mathcal{B}$ can be decomposed as a linear combination of the product of the diagonal elements $\prod_{j=1}^n\langle \mathfrak{m}'_i(k)| \mA_j |\mathfrak{m}'_i(k)\rangle$ in the basis $\mathcal{B}^k$.

The sum rule (\ref{sumrulegeneral}) gives the relation between two different concepts of entropy.
Entanglement entropy as well as R\'enyi entropy are important measures widely employed to characterize quantum entanglement between two given subsystems, and its significance has been extensively discussed from various perspectives, ranging from quantum many-body systems to QFTs and quantum gravity \cite{Amico:2007ag, Eisert:2008ur, Calabrese:2009qy, Rangamani:2016dms}. In the context of AdS/CFT correspondence \cite{Maldacena:1997re, Gubser:1998bc, Witten:1998qj}, entanglement entropy is found to be associated with the minimal surface in the bulk \cite{Ryu:2006bv, Hubeny:2007xt}, known as the Ryu-Takayanagi (RT) formula and its generalization Hubeny-Rangamani-Takayanagi (HRT) formula.
This discovery opens up a new pathway for understanding the quantum nature of black holes and general spacetime \cite{VanRaamsdonk:2010pw, Almheiri:2014lwa, Penington:2019npb, Almheiri:2019psf}.

The concept of pseudoentropy is quite new.  The main drive behind exploring pseudoentropy lies in its capacity to satisfy the RT formula in a new class of non-Hermitian state, providing valuable insights into geometry and entanglement \cite{Nakata:2020luh}. Additionally, it serves as a novel order parameter in quantum many-body systems, extending beyond the entanglement entropy \cite{Mollabashi:2020yie, Kitaev:2005dm, Levin:2005vvn}.
Unlike conventional entropy, pseudoentropy is generally not a real number.
It has been revealed in \cite{Guo:2022jzs} that the reality condition of pseudoentropy is closely associated with non-Hermitian physics \cite{Ashida:2020dkc, Bender:2007nj, Mostafazadeh:2008pw}. Moreover, the imaginary part of pseudoentropy has been proposed in \cite{Doi:2022iyj} to be linked to time-like entanglement and the emergence of time in the dS/CFT correspondence.
Overall, pseudoentropy exhibits significant potential for applications across various aspects of physics.
For more recent developments, one can refer to \cite{Mollabashi:2021xsd, Nishioka:2021cxe, Goto:2021kln, Miyaji:2021lcq, Akal:2021dqt, Guo:2022sfl, Mukherjee:2022jac, Ishiyama:2022odv, Narayan:2022afv, Li:2022tsv, He:2023eap, Doi:2023zaf, Jiang:2023ffu, Narayan:2023ebn, He:2023wko}.

Even though many interesting results have been obtained, the physical significance and potential applications of transition matrix as well as pseudoentropy are still far from clear.
The sum rules we have obtained imply that they are not merely a simple generalization, but are closely related to the concept of entanglement. These researches also indicate that certain classes of transition matrices may possess a dual representation in terms of bulk geometry. However, our current understanding of this correspondence is still limited. The sum rule (\ref{sumrulegeneral}) establishes a link between the geometry associated with the transition matrix and the density matrix, shedding light on this intriguing relationship.

The sum rules are crucial for our understanding of the transition matrix, pseudoentropy, and their holographic dual. In the following sections, we will briefly discuss possible applications of the sum rules in these aspects.

\section{Proof of the operator sum rule}\label{sectionPro}

According to the definition, we have
\bea
&& \rho_A(c) = \mathcal{N}(c)^2 \big( \rho_A^\phi+c \braphi \psi \rangle\transA \nn\\
&& \phantom{\rho_A(c) =}
              + c^* \brapsi \phi\rangle \transAc+c c^* \rho_A^\psi \big),
\eea
where $\rho_A^\phi:= \tr_{\bar A} \ketphi \braphi$, and $\rho_A^\psi:= \tr_{\bar A} \ketpsi \brapsi$.
The $n$-th power of $\rho_A(c)$ can be expanded as
\bea
&& \hspace{-6mm} [\rho_A(c)]^n=\sum_{\{r,s,t\}}c^{t+r} (c^*)^{s+r}\braphi \psi \rangle^t \brapsi \phi\rangle^s\nn \mathcal{N}(c)^{2n}\\
&& \hspace{-18mm} \phantom{[\rho_A(c)]^n=} \times \{(\rho_A^\phi)^{n-r-s-t} (\transA)^t (\transAc)^s (\rho^\psi_A)^{r} + \cdots \},
\eea
where $``+ \cdots "$ denotes the sum of all possible permutation terms for each fixed $r,s,t$.
For example, when $n=5$, $r=4$, $t=0$ and $s=1$, we have the expression $\{(\rho_A^\phi)^{4} \transA + \cdots \} = (\rho_A^\phi)^{4} \transA +(\rho_A^\phi)^{3} \transA \rho_A^\phi+(\rho_A^\phi)^{2} \transA (\rho_A^\phi)^{2}+\rho_A^\phi \transA (\rho_A^\phi)^{3}+ \transA (\rho_A^\phi)^{4}$.
For a fixed set $\{r,s,t\}$, there are $\frac{n!}{r! s! t! (n-r-s-t)!}$ terms.
Now, consider a linear combination of $[\rho_A(c)]^n$
\bea
&& \hspace{-5mm} \sum_{c\in \mathcal{S}} a(c)  [\rho_A(c)]^n =\sum_{\{r,s,t\}} \sum_{c\in \mathcal{A}} a(c) c^{t+r} (c^*)^{s+r}\braphi \psi \rangle^t \brapsi \phi\rangle^s \nn \\
&& \hspace{-3mm} \times\mathcal{N}(c)^{2n}\{(\rho_A^\phi)^{n-r-s-t} (\transA)^t (\transAc)^s (\rho^\psi_A)^{r} + \cdots \}.
\eea
If there is the condition
\be\label{constraint1}
\sum_{c\in \mathcal{S}} a(c) c^{t+r} (c^*)^{s+r}\braphi \psi \rangle^t \brapsi \phi\rangle^s\mathcal{N}(c)^n  =\delta_{s,0}\delta_{t,n},
\ee
the operator sum rule (\ref{operatorsumrule}) is obtained.

To find a solution to (\ref{constraint1}), we consider the set $\mathcal{S}=\big\{\ep^{\frac{2\pi \ii}{2n+1}k} \big| k=0,1,2,\cdots,2n\big\}$.
Equation (\ref{constraint1}) can be solved as
\bea\label{constraint1solve}
 \sum_{k=0}^{2n} f_k  \ep^{-\frac{2\pi \ii}{2n+1}k(n+s-t)}=\delta_{s,0}\delta_{t,n},
 \eea
where
\bea\label{fkak}
f_k:=a(\ep^{\frac{2\pi \ii}{2n+1}k})\left[\ep^{\frac{2\pi \ii}{2n+1}k } \mathcal{N}(\ep^{\frac{2\pi \ii}{2n+1}k})^2\braphi \psi\rangle\right]^n.
\eea
The value of $f_k$ can be found using an inverse discrete Fourier transform, resulting in $f_k=\frac{1}{2n+1}$. We obtain $a_k:= a(\ep^{\frac{2\pi \ii}{2n+1}k})$ (\ref{ak}) using the relation (\ref{fkak}).

By replacing the right-hand side of (\ref{constraint1}) as well as (\ref{constraint1solve}) with $\delta_{s,n}\delta_{t,0}$, we arrive at the sum rule involving $\mathcal{T}_A^{\phi|\psi}$. Similarly, we can find a relation for the transition matrix $\mathcal{T}_A^{\phi|\psi}$
\bea
\ep^{(1-n)S^{(n)}(\mathcal{T}_A^{\phi|\psi})}=\sum_{k=0}^{2n}\ep^{\frac{2\pi \ii}{2n+1}k n} \frac{\ep^{(1-n)S^{(n)}(\rho_A(k))}}{(2n+1)\mathcal{N}_k^n}.
\eea
This is consistent with the fact that $[S^{(n)}(\transA)]^*=S^{(n)}(\cT_A^{\phi|\psi})$. The transition matrix can also be defined for a pair of orthogonal states.
If the two states $\ketphi$ and $\ketpsi$ are orthogonal, the transition matrix is modified to $\trans=|\psi\rangle \langle \phi|$. The reduced transition matrix is defined as $\transA= \tr_{\bar A} |\psi\rangle \langle \phi|$.
The sum rule is slightly different in this case. The normalization constant becomes $\mathcal{N}(c)=1/\sqrt{1+|c|^2}$. To obtain the sum rule we require the similar relation as
\bea\label{SMconstraintorth}
\sum_{c\in \mathcal{S}}a'(c) [\mathcal{N}(c)]^n (c^*)^s c^t=\delta_{s,0}\delta_{t,n}.
\eea
We still take $\mathcal{S}=\big\{\ep^{\frac{2\pi \ii}{2n+1}k} \big| k=0,1,2,\cdots,2n\big\}$. It is not hard to get
\bea
a'_k:=a'(\ep^{\frac{2\pi \ii}{2n+1}k})= \frac{2^n}{2n+1}\ep^{-\frac{2\pi \ii}{2n+1}kn}.
\eea
The sum rule for orthogonal states are given by
\be\label{SMsumrulefororth}
\tr_A(\transA)^n=\sum_{k=0}^{2n} a'_k \tr_A [\rho_A(k)]^n,
\ee
where $\rho_A(k):= \tr_{\bar A} |\xi(\ep^{\frac{2\pi \ii}{2n+1}k})\rangle \langle \xi(\ep^{\frac{2\pi \ii}{2n+1}k})| $.


\section{Sum rule for pseudo-R\'enyi entropy from replica method in QFTs}\label{sectionQFT}

In quantum field theory (QFT), the computation of (pseudo) R\'enyi entropy can be achieved using the replica method \cite{Calabrese:2004eu,Nakata:2020luh}. Additionally, one can establish the formula presented in (\ref{sumrulegeneral}) using the same replica method. The state $\ketpsi$ and $\ketphi$, as well as their superposition $|\xi(c)\rangle=\mathcal{N}(c)(\ketphi+c \ketpsi)$ can be prepared using Euclidean path-integral with operator insertions. The replica partition function in the transition matrix $\trans$ can be computed by path integral on an $n$-sheet manifold ${\mathcal{R}}_n$ gluing along subsystem $A$,
\be\label{SMpseudocorrelation}
\tr_A [(\transA)^n]=\frac{\mathcal{Z}_n}{\braphi \psi\rangle^n} \Big\langle \prod^n_{i=1} \phi(i)^\dagger \prod^n_{j=1} \psi(j)\Big\rangle_{\mathcal{R}_n},
\ee
where $\mathcal{Z}_n$ is the vacuum partition function on $\mathcal{R}_n$, $\psi(j)$ and $\phi(i)$ denote the operators on the $j$-th and the $i$-th sheets, respectively. Similarly, for the reduced density matrix $\rho_A(c)$ we have
\bea\label{SMtracerhoc}
&&\tr_A \{ [\rho_A(c)]^n \} = \mathcal{Z}_n \mathcal{N}(c)^n
                        \Big\langle \prod^n_{i=1} [ \phi(i)^\dagger+c^* \psi(i)^\dagger ] \nn \\
&&\phantom{\tr_A \{ [\rho_A(c)]^n \} =}
                        \times\prod^n_{j=1} [ \phi(j)+c\psi(j) ] \Big\rangle_{\mathcal{R}_n}.
\eea
The correlation function in the above equation can be expanded as
\bea \label{SMexpand}
&& \phantom{=} \Big \langle \prod^n_{i=1} [ \phi(i)^\dagger+c^* \psi(i)^\dagger ]
                            \prod^n_{j=1} [ \phi(j)+c\psi(j) ] \Big\rangle_{\mathcal{R}_n}\nn \\
&& = \sum_{\{s,t\}} (c^*)^s c^t\langle  \psi(i_1)^\dagger \cdots \psi(i_s)^\dagger \phi(i_1')^\dagger\cdots\phi(i'_{n-s})^\dagger \nn\\
&& \phantom{=}
     \times \psi(j_1)\cdots\psi(j_t) \phi(j_1')\cdots\phi(j'_{n-t})\rangle_{\mathcal{R}_n},
\eea
where $s,t \in (0,1,...,n)$. For fixed $s,t$ there are $\frac{(n!)^2}{s!t!(n-s)!(n-t)!}$ terms, thus the total number of terms in the sum is $\sum_{s,t=0}^{n}\frac{(n!)^2}{s!t!(n-s)!(n-t)!}=2^{2n}$. One of the special term is the one with $s=0,t=n$, i.e., $\langle \prod^n_{i=1} \phi(i)^\dagger \prod^n_{j=1} \psi(j)\rangle_{\mathcal{R}_n}$, which is the correlation function in (\ref{SMpseudocorrelation}).

Then we consider linear combinations of $\tr_A[\rho_A(c)]^n$ (\ref{SMtracerhoc}). By choosing a series of functions $a(c)$ with the index $c\in \mathcal{S}$, we have
\bea
&& \phantom{=} \sum_{c\in \mathcal{S}} a(c)  \tr_A [\rho_A(c)]^n \\
&&=\mathcal{Z}_n\sum_{\{s,t\}}\sum_{c\in \mathcal{A}}a(c) [ \mathcal{N}(c)]^n (c^*)^s c^t \big\langle  \psi(i_1)^\dagger...\psi(i_s)^\dagger \nn\\
&& \phantom{=} \times \phi(i_1')^\dagger...\phi(i'_{n-s})^\dagger  \psi(j_1)...\psi(j_t) \phi(j_1')...\phi(j'_{n-t}) \big\rangle_{\mathcal{R}_n}.\nn
\eea
If we have
\be
\sum_{c\in \mathcal{S}}a(c)[\mathcal{N}(c)]^n (c^*)^s c^t=\frac{1}{\braphi \psi\rangle^n}\delta_{s,0}\delta_{t,n},
\ee
the above summation would reduce to $\tr_A(\transA)^n$, which is just the sum rule (\ref{operatorsumrule}). To find a solution, one could take $\mathcal{S}=\big\{\ep^{\frac{2\pi \ii}{2n+1}k} \big| k=0,1,2,\cdots,2n\big\}$, we would have the same equations (\ref{constraint1solve})(\ref{fkak}). Thus we have proved the sum rule (\ref{opsummain}) by the replica method.

\section{Short interval expansion of pseudo-R\'enyi entropy in CFTs} \label{sectionCFT}

Another approach to evaluate the (pseudo) R\'enyi entropy involves utilizing the twist operator formalism. In certain special cases, like a short interval in 2-dimensional conformal field theories (CFTs), the operator product expansion (OPE) of the twist operator can be employed \cite{Cardy:2007mb,Headrick:2010zt,Calabrese:2010he,Rajabpour:2011pt,Chen:2013kpa}. By doing so, the sum rule (\ref{sumrulegeneral}) is expressed as a series of relationships between correlation functions of local operators, thus we can derive the relations (\ref{correlator_sumrule_nonorth}) and (\ref{correlator_sumrule_orth}).

The sum rule shows the $\tr_A(\transA)^n$ can be written as linear combination of $\tr_A(\rho_A(c))^n$.  The R\'enyi entropy can be evaluated by using twist operator formalism in certain cases \cite{Calabrese:2004eu}. An obvious conclusion from the sum rule is that pseudo-R\'enyi entropy can also be evaluated by twist operator formalism in same case.
In this section we focus on two-dimensional CFTs. Consider the CFT is defined on a cylinder with spatial period $L$. For an interval $A=[0,\ell]$ with $\ell \ll L$, one could use operator product expansion of twist operator \cite{Cardy:2007mb,Headrick:2010zt,Calabrese:2010he,Rajabpour:2011pt,Chen:2013kpa}.
Generally for a given reduced density matrix $\rho_A$, there is
\bea\label{SMshort}
&& \tr_A(\rho_A^n)=\left(\frac{\epsilon}{\ell}\right)^{4h_\sigma}
                   \Big[ 1+\sum_{m=1}^n \sum_{\mathcal{X}_1,...,\mathcal{X}_m} 	\ell^{\Delta_{\mathcal{X}_1}+...+\Delta_{\mathcal{X}_m}}\nn \\
&& \phantom{\tr_A(\rho_A^n)=} \times d_{\mathcal{X}_1...\mathcal{X}_m} \tr(\rho \mathcal{X}_1) \cdots \tr(\rho \mathcal{X}_m) \Big],
\eea
where the summation $\{\mathcal{X}_1,\cdots,\mathcal{X}_m \}$ is over all the quasi-primary operators, $\Delta_{\mathcal{X}_i}$ are their scaling dimension, $h_\sigma$ is the conformal weight of twist operator, $d_{\mathcal{X}_1...\mathcal{X}_m}$ are the coefficients independent with the state $\rho_A$. For the transition matrix $\transA$, the formula is almost the same
\bea\label{SMshortpseudo}
&& \tr_A(\transA)^n = \left(\frac{\epsilon}{\ell}\right)^{4h_\sigma} \Big[ 1+\sum_{m=1}^n \sum_{\mathcal{X}_1,...,\mathcal{X}_m} 	\ell^{\Delta_{\mathcal{X}_1}+...+\Delta_{\mathcal{X}_m}} \nn \\
&& \hspace{10mm} \times d_{\mathcal{X}_1 \cdots \mathcal{X}_m} \tr(\trans \mathcal{X}_1) \cdots \tr(\trans \mathcal{X}_m) \Big].
\eea
Defining $F_{m}(c):= \tr(\rho(c)\mathcal{X}_1)\cdots\tr(\rho(c)\mathcal{X}_m)$, we have
\bea\label{SMequationFm}
&& F_m(\ep^{\frac{2\pi \ii}{2n+1}k})
   = [ \mathcal{N}(\ep^{\frac{2\pi \ii}{2n+1}k}) ]^{2m} \prod_{j=1}^m \big( \braphi \mathcal{X}_j \ketphi \\
&& \hspace{4mm} +\ep^{\frac{2\pi \ii}{2n+1}k} \braphi \mathcal{X}_j \ketpsi
   +\ep^{-\frac{2\pi \ii}{2n+1}k} \brapsi \mathcal{X}_j \ketphi+\brapsi \mathcal{X}_j \ketpsi \big).\nn
\eea
The sum rule is given by using the following equation
\be\label{SMequationfourier}
\sum_{k=0}^{2n}\ep^{-\frac{2\pi \ii}{2n+1}kn} \frac{F_{m}(\ep^{\frac{2\pi \ii}{2n+1}k})}{(2n+1)\mathcal{N}_k^n}= \frac{1}{\braphi \psi\rangle^m}\prod_{j=1}^m\braphi \mathcal{X}_j \ketpsi,
\ee
which can be shown directly by using (\ref{SMequationFm}).
Note that
\be
\mathcal{N}(\ep^{\frac{2\pi \ii}{2n+1}k})=\big(2+\ep^{\frac{2\pi \ii}{2n+1}k} \braphi \psi\rangle +\ep^{-\frac{2\pi \ii}{2n+1}k} \brapsi \phi\rangle\big)^{-1/2}.
\ee
The summation over $k$ is discrete Fourier transform of the terms in the square brackets. These terms are series in  $\ep^{\frac{2\pi \ii}{2n+1}k}$. The coefficient corresponding to the highest power  $\ep^{\frac{2\pi \ii}{2n+1}kn}$ is
\be
\brapsi \phi\rangle^{n-m}\prod_{j=1}^m\braphi \mathcal{X}_j \ketpsi.
\ee
In fact this is the only term survived in the discrete Fourier transform. Obviously, (\ref{SMequationfourier}) holds.

In fact, Equation (\ref{SMequationfourier}) represents the sum rule for the off-diagonal matrix elements (\ref{correlator_sumrule_nonorth}). The relation (\ref{correlator_sumrule_orth}) can be derived by applying the operator sum rule to the orthogonal states (\ref{SMsumrulefororth}).

\section{Examples for the sum rule}\label{sectionSim}

\subsection{Two qubits system}

The sum rule can be examined in a system of two qubits represented in the basis states $|00\rangle, |01\rangle, |10\rangle, |11\rangle$.
We assume two given states
\bea
&&|\phi\rangle= \frac{1}{\sqrt{ | \alpha_\phi|^2  + |\beta_\phi|^2}}(\alpha_\phi|00\rangle +\beta_\phi|11\rangle),\nn\\
&&|\psi\rangle= \frac{1}{\sqrt{ | \alpha_\psi|^2  + |\beta_\psi|^2}}(\alpha_\psi|00\rangle +\beta_\psi|11\rangle),
\eea
where $\alpha_{\phi(\psi)}$ and $\beta_{\phi(\psi)}$ represent arbitrary constants.
The superposition state is  $|\xi(k)\rangle:= \mathcal{N}(\ep^{\frac{2\pi \ii}{2n+1}k})(\ketphi+\ep^{\frac{2\pi \ii}{2n+1}k})\ketpsi$.

The reduced density matrix $\rho_A(k)$ is given by
\begin{eqnarray}
\rho_A(k)=
\begin{pmatrix}
\lambda_1 & 0 \\
0 & \lambda_2
\end{pmatrix},
\end{eqnarray}
with
\bea
&& \lambda_1 = [\mathcal{N}(\ep^{\frac{2\pi \ii k}{2n+1}})]^2
               \Big| \frac{\alpha _{\phi }}
                          {\sqrt{ |\alpha_{\phi}|^2 +|\beta_{\phi}|^2}}
                  + \frac{\alpha _{\psi } \ep^{\frac{2 \ii \pi  k}{2 n+1}}}
                         {\sqrt{ |\alpha_{\psi}|^2 +|\beta_{\psi}|^2}}
               \Big|^2, \nn\\
&& \lambda_2 = [\mathcal{N}(\ep^{\frac{2\pi \ii k}{2n+1}})]^2
               \Big| \frac{\beta _{\phi }}
                          {\sqrt{ |\alpha_{\phi}|^2 +|\beta_{\phi}|^2}}
                    +\frac{\beta _{\psi } \ep^{\frac{2 \ii \pi  k}{2 n+1}}}
                          {\sqrt{ |\alpha_{\psi}|^2 +|\beta_{\psi}|^2}}\Big|^2, \nn
\eea
where
\bea
&& [\mathcal{N}(\ep^{\frac{2\pi \ii k}{2n+1}})]^{-2}
  =\Big| \frac{\alpha _{\phi }}
                          {\sqrt{ |\alpha_{\phi}|^2 +|\beta_{\phi}|^2}}
                  + \frac{\alpha _{\psi } \ep^{\frac{2 \ii \pi  k}{2 n+1}}}
                         {\sqrt{ |\alpha_{\psi}|^2 +|\beta_{\psi}|^2}}
               \Big|^2 \nn\\
&& \hspace{8mm}
   + \Big| \frac{\beta _{\phi }}
                          {\sqrt{ |\alpha_{\phi}|^2 +|\beta_{\phi}|^2}}
                    +\frac{\beta _{\psi } \ep^{\frac{2 \ii \pi  k}{2 n+1}}}
                          {\sqrt{ |\alpha_{\psi}|^2 +|\beta_{\psi}|^2}}\Big|^2.
\eea
The reduced transition matrix $\transA$ is given by
\bea
\transA=\left(
\begin{array}{cc}
 t_1 & 0 \\
 0 & t_2 \\
\end{array}
\right),
\eea
with
\be
t_1=\frac{\a^*_\phi \alpha _{\psi }}{\a^*_\phi \alpha _{\psi }+\b^*_\phi \beta _{\psi }},
~~
t_2=\frac{\b^*_\phi \beta _{\psi }}{\a^*_\phi \alpha _{\psi }+\b^*_\phi \beta _{\psi }}.\nn
\ee
To check the sum rule (\ref{opsummain}) one just needs
\be
\frac{1}{(2n+1) \braphi \psi\rangle^n}\sum_{k=0}^{2n}\ep^{-\frac{2\pi \ii }{2n+1}kn} \frac{\lambda_i^n}{\mathcal{N}(\ep^{\frac{2\pi \ii}{2n+1}})^{2n}}=t_i^n,
\ee
with $i=1,2$, where
\be
\braphi \psi\rangle = \frac{\alpha _{\psi } \a^*_\phi+\beta _{\psi } \b^*_\phi}
                           { \sqrt{\alpha _{\psi } \a^*_\psi+\beta _{\psi } \b^*_\psi}
                             \sqrt{\alpha _{\phi } \a^*_\phi+\beta _{\phi } \b^*_\phi} }.
\ee

\subsection{Perturbation states}

We examine the states $\ketpsi$ and $\ketphi=\mathcal{N}_\phi(|\psi\rangle +\epsilon |\psi'\rangle)$ with $\mathcal{N}_\phi=1-\frac{1}{2}(\epsilon \langle \psi | \psi'\rangle +\epsilon^* \langle \psi' | \psi\rangle)+O(\epsilon^2)$, considering only to the leading order of $\epsilon$. The reduced transition matrix is then expressed as
\be
\transA=\mathcal{N}(\epsilon)(\rho_A^\psi+ \epsilon^* \mathcal{T}_{A}^{\psi| \psi'}),
\ee
where $\mathcal{N}(\epsilon)=1-\epsilon^* \langle \psi'| \psi\rangle$, $\rho_A^\psi= \tr_{\bar A}|\psi\rangle \langle \psi|$, and $\mathcal{T}_{A}^{\psi| \psi'}=\tr_{\bar A}|\psi\rangle \langle \psi'|$.
We can then write the expression for $\tr_A(\transA)^n$ as
\bea\label{perturbtrans}
&& \tr_A(\transA)^n=\tr_A(\rho_A^\psi)^n
                  \Big(   1-n \epsilon^* \langle \psi'| \psi\rangle \\
&& \phantom{\tr_A(\transA)^n=}
                        + n \epsilon^*
                        \frac{\tr_A[(\rho_A^\psi)^{n-1}\mathcal{T}_A^{\psi|\psi'}]}{\tr_A(\rho_A^\psi)^n} \Big).\nn
\eea
The superposition state is given by
\be
|\xi(k)\rangle = \mathcal{N}(\ep^{\frac{2\pi \ii}{2n+1}k})
                 \Big[ (\mathcal{N}_\phi+\ep^{\frac{2\pi \ii}{2n+1}k})|\psi\rangle +\epsilon |\psi'\rangle \Big],
\ee
where $\mathcal{N}(\ep^{\frac{2\pi \ii}{2n+1}k})$ is the normalization constant.
The expression for $\tr_A[\rho_A(k)]^n$ is then
\ble
\tr_A[\rho_A(k)]^n=(\tilde{\mathcal{N}}_k)^{-n} \tr_A(\rho_A^\psi)^n
\Big[ 1
   + \frac{n \epsilon^*}{1+\ep^{-\frac{2\pi \ii}{2n+1}k}}
     \frac{\tr_A[(\rho_A^\psi)^{n-1}\mathcal{T}_A^{\psi|\psi' }]}{\tr_A (\rho_A^\psi)^n }
   + \frac{n\epsilon}{1+\ep^{\frac{2\pi \ii}{2n+1}k}}
     \frac{\tr_A[(\rho_A^\psi)^{n-1}\mathcal{T}_A^{\psi'|\psi}]}{\tr_A (\rho_A^\psi)^n }
\Big],
\ele
where $\tilde{\mathcal{N}}_k=1+\frac{\epsilon \langle \psi |\psi'\rangle }{1+\ep^{\frac{2\pi \ii}{2n+1}k}}+\frac{\epsilon^* \langle \psi' |\psi\rangle}{1+\ep^{-\frac{2\pi \ii}{2n+1}k}}$.

For the superposition state
\be
|\xi(k)\rangle= \mathcal{N}(\ep^{\frac{2\pi \ii}{2n+1}k})\left[(\mathcal{N}_\phi+\ep^{\frac{2\pi \ii}{2n+1}k})|\psi\rangle +\epsilon |\psi'\rangle\right],
\ee
the normalization constant is
\bea
&& \hspace{-8mm} \mathcal{N}(\ep^{\frac{2\pi \ii}{2n+1}k})= |1+\ep^{\frac{2\pi \ii}{2n+1}k}|^{-1} \Big( 1-\frac{1}{4}\epsilon \langle \psi |\psi'\rangle \frac{\ep^{-\frac{2\pi \ii}{2n+1}k}-1}{\ep^{-\frac{2\pi \ii}{2n+1}k}+1}\nn \\
&& \hspace{-8mm} \phantom{\mathcal{N}( \ep^{\frac{2\pi \ii}{2n+1}k})= }-\frac{1}{4}\epsilon^* \langle \psi |\psi'\rangle \frac{\ep^{\frac{2\pi \ii}{2n+1}k}-1}{\ep^{\frac{2\pi \ii}{2n+1}k}+1} \Big).
\eea
We also have
\bea
\langle \phi |\psi\rangle= 1+\frac{1}{2}\epsilon \langle \psi |\psi'\rangle-\frac{1}{2}\epsilon^* \langle \psi' |\psi\rangle.
\eea
It is straightforward to obtain
\bea\label{SMperturb1}
&& \phantom{=}
   \sum_{k=0}^{2n}\ep^{-\frac{2\pi \ii}{2n+1}k n} \frac{\tr_A\rho_A(k)^n}{(2n+1)\mathcal{N}(\ep^{\frac{2\pi \ii}{2n+1}k})^{2n}\langle \phi |\psi\rangle^n}\nn \\
&& =\frac{1}{2n+1}\sum_{k=0}^{2n}
 \ep^{-\frac{2\pi \ii}{2n+1}k n}p_k,
\eea
where
\ble
p_k= |1+\ep^{\frac{2\pi \ii}{2n+1}k}|^{2n}
\tr_A(\rho_A^\psi)^n \Big( 1-n\epsilon^*\langle \psi'| \psi\rangle
+ \frac{n\epsilon }{1+\ep^{\frac{2\pi \ii}{2n+1}k}}\frac{\tr_A[(\rho_A^\psi)^{n-1}\mathcal{T}_A^{\psi'|\psi}]}{\tr_A (\rho_A^\psi)^n }
+ \frac{n \epsilon^*  }{1+\ep^{-\frac{2\pi \ii}{2n+1}k}}\frac{\tr_A[(\rho_A^\psi)^{n-1}\mathcal{T}_A^{\psi'|\psi}]}{\tr_A (\rho_A^\psi)^n } \Big).
\ele
The non-vanishing terms in the summation (\ref{SMperturb1}) are the ones associated with $\ep^{\frac{2\pi \ii}{2n+1}kn}$, which is equal to Eq.~(\ref{perturbtrans}). Therefore, the sum rule is proved in this example.\\

\subsection{Locally excited states in two-dimensional massless boson theory}\label{sectionquasi}

For a general transition matrix
\bea\label{SMtransionlocal}
\trans=\frac{\mO(w_1,\bar w_1)|0\rangle\langle 0| \mtO^\dagger(w_2,\bar w_2)}{\langle 0| \mtO^\dagger(w_2,\bar w_2) \mO(w_1,\bar w_1)|0\rangle},
\eea
one could use replica method to evaluate the (pseudo) R\'enyi entropy.
The two states are $\ketphi=\td\mO(w_2,\bar w_2)|0\rangle$ and $\ketpsi=\mO(w_1,\bar w_1)|0\rangle$.
We focus on the pseudo-R\'enyi entropy with $n=2$. It can be shown that the pseudo-R\'enyi entropy is associated with the four point functions on the 2-sheet Riemann surface $\Sigma_2$
\bea
&& \phantom{=} \Delta S^{(2)}(\transA) \\
&& = -\log
\frac{\langle\mO(w_1,\bar w_1)\mtO^\dagger(w_2,\bar w_2)\langle\mO(w_3,\bar w_3)\mtO^\dagger(w_4,\bar w_4)\rangle_{\Sigma_2}  }
     {\langle  \mtO^\dagger(w_2,\bar w_2) \mO(w_1,\bar w_1)\rangle_{\Sigma_1}},\nn
\eea
with $\Delta S^{(2)}(\transA):= S^{(2)}(\transA)-S^{(2)}(\rho_A^{(0)})$, where $\rho_A^{(0)}$  is the reduced density matrix for ground state, $w_i\in \Sigma_{2}$ ($i=1,2,3,4$) are coordinates of the corresponding local operators.

To calculate the four point correlation function on $\Sigma_2$, one could use the conformal mapping
\bea
z=\sqrt{\frac{w-x_1}{w-x_2}},
\eea
where $x_1,x_2$ are endpoints of the interval $A$. The corresponding coordinates of $w_i \in \Sigma_2$ under the conformal mapping are given by
\be
z_1=-z_3=\sqrt{\frac{w_1-x_1}{w_1-x_2}}, ~~
z_2=-z_4=\sqrt{\frac{w_2-x_1}{w_2-x_2}}.
\ee
We take $w=x+\ii\tau$, $\bar w=x-\ii \tau$. We mainly focus on the operators
\bea
&& \mO(w_1,\bar w_1)= :\ep^{\frac{\ii}{2}\phi(0,-  a)}:+\ep^{\ii\theta}:\ep^{-\frac{\ii}{2}\phi(0,- a)}:, \nn \\
&& \mtO^\dagger(w_2,\bar w_2)=:\ep^{\frac{\ii}{2}\phi(0,  a')}:+:\ep^{-\frac{\ii}{2}\phi(0, a')}:,
\eea
with $w_1=-\ii a$ and $w_2=\ii a'$. The two states under consideration are
\bea\label{2CFTstates}
&&|\phi\rangle= (\ep^{\frac{\ii}{2}\phi(0,-a)}+\ep^{-\frac{\ii}{2}\phi(0,- a)})|0\rangle,\nn \\
&&|\psi\rangle= (\ep^{\frac{\ii}{2}\phi(0,a')}+\ep^{\ii \theta-\frac{\ii}{2}\phi(0, a')})|0\rangle.
\eea
$\Delta S^{(2)}(\transA)$ has been evaluated in \cite{Nakata:2020luh}, and the result is
\bea
\Delta S^{(2)}(\transA)=\log \frac{1+\cos\theta}{\cos\theta +|\eta|^2+|1-\eta|^2},
\eea
where the cross ratio $\eta=\frac{(z_1-z_2)(z_3-z_4)}{(z_1-z_3)(z_2-z_4)}$. For the case $\theta=0$, the pseudo-R\'enyi entropy reduces to R\'enyi entropy. In the limit $a,a'\to 0$, $\eta,\bar \eta \to 0$, $\Delta S^{(2)}(\transA)\to 0$. For the case $\theta \ne 0$, we can also consider the Lorentzian time evolution by analytic continuation $a\to \delta +it $ and $a' \to \delta -i t$ where $\delta$ is the UV cutoff.  It can be shown that in the region $x_1<t<x_2$ we will have
\bea
\Delta S^{(2)}(\transA)=\log \frac{1+\cos \theta}{\cos\theta}.
\eea

We take $a=a'$. The superposition state $|\xi(k)\rangle$ is given by
\bea\label{SMsuperEPR}
|\xi(k)\rangle =\mathcal{N}(\ep^{\frac{2\pi \ii}{2n+1}k})(\alpha_k :\ep^{\frac{\ii}{2}\phi(0,-a)}:+\beta_k :\ep^{-\frac{\ii}{2}\phi(0,- a)}: )|0\rangle,\nn\\
\eea
where
\bea
\alpha_k =1+\ep^{\frac{2\pi \ii}{2n+1}k},\quad \beta_k=1+\ep^{\ii\theta+\frac{2\pi \ii}{2n+1}k}.
\eea
The R\'enyi entropy $\Delta S^{(2)}(\rho_A(k)):=S^{(2)}(\rho_A(k))-S^{(2)}(\rho^{(0)}_A)$ can also be evaluated by the replica method. With some calculations we get
\bea
&& \phantom{=} \Delta S^{(2)}(\rho_A(k)) \\
&&=\log \frac{(|\alpha_k |^2+|\beta_k|^2)^2}{|\alpha_k |^4+2|\alpha_k |^2|\beta_k|^2 |\eta|+2|\alpha_k |^2|\beta_k|^2 |1-\eta|+  |\beta_k|^4}.\nn
\eea
For the case $\theta=0$ and $a\to 0$ we have $\Delta S^{(2)}(\rho_A(k))\to 0$, which is consistent with \cite{He:2014mwa}. For the case $\theta\ne 0$, let us consider the Lorentzian time evolution by analytic continuation $a\to \delta +it $. In the region $x_1<t<x_2$ we have $\eta=1,\bar \eta\to 0$, thus
\bea
\Delta S^{(2)}(\rho_A(k))=\log(\frac{(|\alpha_k |^2+|\beta_k|^2)^2}{|a_k|^4+|\beta_k|^4}).
\eea
One could check $\Delta S^{(2)}(\transA)$ and $\Delta S^{(2)}(\rho_A(k))$ satisfy the sum rule (\ref{sumrulegeneral}) with $n=2$. Further, the $n$-th (pseudo) R\'enyi entropy can be evaluated by replica method. In section.\ref{sectionquasiparticle}, we will elucidate the aforementioned simple formula using the quasiparticle framework, thereby providing an explanation of quasiparticles for the pseudo R\'enyi entropy.

\subsection{Quasiparticle excited states in spin chain}

We consider a chain of $L$ free bosons or fermions with Hamiltonian
\be
H = \sum_{j=1}^L a_j^\dag a_j.
\ee
The global modes are
\be
b_k = \f{1}{\sr{L}} \sum_{j=1}^L \ep^{-\f{2\pi\ii j k}{L}} a_j, ~~
b_k^\dag = \f{1}{\sr{L}} \sum_{j=1}^L \ep^{-\f{2\pi\ii j k}{L}} a_j.
\ee
The ground state $|G\rag$ is defined as
\be
a_j|G\rag = 0, ~ \forall j.
\ee
For the subsystem $A=[1,\ell]$ and its complement $B=[\ell+1,L]$, the ground state is a direct product state
\be
|G\rag = |G_A\rag \otimes |G_B\rag,
\ee
where the subsystem ground states are defined as
\bea
&& a_j |G_A\rag = 0, ~ \forall j \in A, \nn\\
&& a_j |G_B\rag = 0, ~ \forall j \in B.
\eea

We will also use the single-quasiparticle excited state
\be
|k\rag = b_k^\dag |G\rag.
\ee
It is convenient to define the subsystem modes
\bea
&& b_{A,k} = \f{1}{\sr{L}} \sum_{j=1}^\ell \ep^{-\f{2\pi\ii j k}{L}} a_j, ~~
   b_{A,k}^\dag = \f{1}{\sr{L}} \sum_{j=1}^\ell \ep^{\f{2\pi\ii j k}{L}} a_j^\dag, \nn\\
&& b_{B,k} = \f{1}{\sr{L}} \sum_{j=\ell+1}^L \ep^{-\f{2\pi\ii j k}{L}} a_j, ~~
   b_{B,k}^\dag = \f{1}{\sr{L}} \sum_{j=\ell+1}^L \ep^{\f{2\pi\ii j k}{L}} a_j^\dag. \nn
\eea

For two different single-quasiparticle excited states $|k_1\rag$ and $|k_2\rag$, which are orthogonal, we may define the reduced transition matrix
\be
\cT_A^{k_2|k_1}=\tr_B|k_2\rag\lag k_1|,
\ee
which could not be normalized.
Then we get the reduced transition matrix
\be
\cT_A^{k_2|k_1} = -\a_{12} |G_A\rag\lag G_A| + b_{A,k_2}^\dag |G_A\rag\lag G_A| b_{A,k_1},
\ee
with
\be
\a_{12} = \ep^{-\f{\pi\ii k_{12} (\ell+1)}{L}} \f{\sin\f{\pi k_{12}\ell}{L}}{L\sin\f{\pi k_{12}}{L}}, ~~ k_{12}=k_1-k_2.
\ee
In the following, we will also use $\a_{21}=\a_{12}^*$.
We get the pseudo-R\'enyi entropy
\be \label{SMtrArAk1k2n}
\tr_A (\cT_A^{k_2|k_1})^n = [1+(-)^n]\a_{12}^n.
\ee

For the state
\be
|\xi(\ep^{\ii\phi})\rag = \f{1}{\sr2} ( |k_1\rag + \ep^{\ii\phi} |k_2\rag ),
\ee
we get the reduced density matrix
\bea
&& \r_{A,\xi(\ep^{\ii\phi})} = \f12 \big\{ [ 2(1-x) - \a_{21}\ep^{-\ii\phi} - \a_{12}\ep^{\ii\phi} ] |G_A\rag\lag G_A| \nn\\
&& \phantom{\r_{A,\xi(\ep^{\ii\phi})} =}
   + \ep^{-\ii\theta} b_{A,k_1}^\dag |G_A\rag\lag G_A| b_{A,k_2} \nn\\
&& \phantom{\r_{A,\xi(\ep^{\ii\phi})} =}
   + \ep^{\ii\theta} b_{A,k_2}^\dag |G_A\rag\lag G_A| b_{A,k_1} \\
&& \phantom{\r_{A,\xi(\ep^{\ii\phi})} =}
   + b_{A,k_1}^\dag |G_A\rag\lag G_A| b_{A,k_1} \nn\\
&& \phantom{\r_{A,\xi(\ep^{\ii\phi})} =}
   + b_{A,k_2}^\dag |G_A\rag\lag G_A| b_{A,k_2} \big\},\nn
\eea
where $x=\f{\ell}{L}$.
The R\'enyi entropy is
\bea \label{SMtrArAxiepiithetan}
&& \tr_A \r_{A,\xi(\ep^{\ii\phi})}^n = \f{1}{2^n} \big\{ [ 2x + \a_{21}\ep^{-\ii\phi} + \a_{12}\ep^{\ii\phi} ]^n\\
&& \phantom{\tr_A \r_{A,\xi(\ep^{\ii\phi})}^n =}
                                    + [ 2(1-x) - \a_{21}\ep^{-\ii\phi} - \a_{12}\ep^{\ii\phi} ]^n \big\}.\nn
\eea
Using (\ref{SMtrArAk1k2n}) and (\ref{SMtrArAxiepiithetan}), we check the sum rule
\be
\tr_A (\cT_A^{k_2|k_1})^n = \f{2^n}{2n+1} \sum_{k=0}^{2n} \ep^{-\f{2\pi\ii k}{2n+1}} \tr_A \r_{A,\xi(\ep^{\f{2\pi\ii k}{2n+1}})}^n.
\ee

\section{Applications} \label{sectionApp}

\subsection{Relation between static quantity and dynamical one}

In quantum mechanics, a sum rule typically represents a connection between a static quantity and the summation over a dynamic one. In this context, we aim to demonstrate that our sum rules exhibit a similar behavior.
Let $|\mathfrak{e}_i\rangle$ ($i=0,1,...$) denote an eigenstate of the Hamiltonian $H$ with energy $E_i$. Given an operator $\mA$ we would like to consider the off-diagonal elements $\langle\mathfrak{e}_i|\mathcal{A}|\mathfrak{e}_j\rangle$. According to (\ref{correlator_sumrule_orth}) we have
\bea\label{static_dynamics}
\langle\mathfrak{e}_i|\mathcal{A}|\mathfrak{e}_j\rangle^n=\sum_{k=1}^{2n} a'_k \langle \mathfrak{s}_{ij}|\mathcal{A}(t_k)|\mathfrak{s}_{ij}\rangle^n,
\eea
where $|\mathfrak{s}_{ij}\rangle:=\frac{1}{\sqrt{2}}(|\mathfrak{e}_i\rangle+|\mathfrak{e}_j\rangle)$, the operator $\mA(t):=e^{i Ht}\mA e^{-i Ht}$, and $t_k=\frac{2\pi k}{(2n+1)(E_i-E_j)}$.

The physical meaning of the above result is clear.
The off-diagonal elements of an operator in energy eigenstates can be associated with the expectation value of this operator at time $t_k$ in the superposition states.
In the special case that the operator $\mA$ is independent with time, that is $\mA=\mA(t)$, as a check of  (\ref{static_dynamics}) we can see the off-diagonal element $\langle\mathfrak{e}_i|\mathcal{A}|\mathfrak{e}_j\rangle$ is vanishing, which can also be derived by using the fact $\langle\mathfrak{e}_i|\mathcal{A}|\mathfrak{e}_j\rangle=e^{i(E_i-E_j)t}\langle\mathfrak{e}_i|\mathcal{A}|\mathfrak{e}_j\rangle$.

The formula may be more useful when the asymptotic behavior of $\mathcal{A}(t)$ is understood in the limit as $t\to \infty$. In the special case where $E_i-E_j$ is significantly smaller than the typical energy of the system, the off-diagonal element $\langle\mathfrak{e}_i|\mathcal{A}|\mathfrak{e}_j\rangle$ can be determined by examining the asymptotic behavior of $\mathcal{A}(t)$.

\subsection{Transition matrix with gravity dual}

The transition matrix extends the potential dual geometry of boundary states to non-Hermitian ones. It is anticipated that only a limited class of transition matrices could be dual to bulk geometry. The fundamental constraint is that the pseudoentropy must be positive.  However, we still do not have a comprehensive understanding of the properties of such a transition matrix.

It is generally expected that the theory dual to gravity should be a gapped large-N CFT \cite{Aharony:1999ti,Harlow:2018fse}. The bulk Newton constant $G$ is proportional to $1/N^2$. If a state $|\Psi\rangle$ in the CFT   can be effectively described by a geometry, the connected two point correlation function of single trace operator $\mathcal{O}_i$ should follow the scaling,
\bea
\langle \Psi| \mathcal{O}_i\mathcal{O}_j|\Psi\rangle-\langle \Psi| \mathcal{O}_i|\Psi\rangle \langle \Psi|\mathcal{O}_j|\Psi\rangle\sim O(N^2).
\eea
More generally, a connected $k$-point function of the rescaled operator $O_i:=\mathcal{O}_i/N$ should be of order $N^{2-k}$. Consequently, in the limit as $N\to \infty$, the correlation functions of $O_i$ exhibit behavior akin to generalized free theory \cite{El-Showk:2011yvt}.

Utilizing (\ref{correlator_sumrule_nonorth}) allows one to demonstrate that if the states $|\xi(k)\rangle$ meet the necessary conditions for state with geometry dual, the transition matrix $\trans$ would also satisfy these conditions. This implies that the sum rule could be employed to construct the geometry dual to the transition matrix, given the geometry dual to $|\xi(k)\rangle$. Furthermore, in holography, the (pseudo) R\'enyi entropy can be linked to the gravitational on-shell action with cosmic brane insertion \cite{Dong:2016fnf, Nakata:2020luh}. The sum rule (\ref{sumrulegeneral}) signifies a connection between the bulk solutions.

\subsection{Quasiparticle pictures for pseudo-R\'enyi entropy}\label{sectionquasiparticle}

In this subsection, we will demonstrate that in certain specific setups with the help of the sum rule (\ref{sumrulegeneral}), the pseudo (R\'enyi) entropy can be evaluated within the quasiparticle framework, akin to the traditional treatment of entanglement entropy. This would provide more insight into its relationship with entanglement.
~\\

\textit{Local excitation in 2-dimensional CFTs.} In section.\ref{sectionquasi} we consider the massless boson CFT in two dimensions with two states defined by (\ref{2CFTstates}). The subsystem $A$ is defined as $[x_1,x_2]$ with $x_2>x_1>0$. In QFTs, the replica method is utilized to evaluate the pseudo-R\'enyi entropy. We will take $a=a'$ in the following. The superposition state $|\xi(k)\rangle$ is given by (\ref{SMsuperEPR}). The state (\ref{SMsuperEPR}) can be effectively taken as entangled left-moving and right-moving quasiparticles. When the right-moving particles enter the subsystem $A$ ($x_1<t<x_2$) the entanglement between $A$ and $\bar A$ will increase by the amount of the entangled pairs.

We explore the real-time evolution by employing analytic continuation, specifically $a\to \delta + \ii t $. If we consider $\delta$ as the UV cut-off, the time evolution of the R\'enyi entropy for the superposition state $|\xi(k)\rangle$ can be understood using a quasiparticle picture. In the region $x_1<t<x_2$, we find that $\eta\to 1$ and $\bar \eta \to 0$, where $\eta$ and $\bar{\eta}$ are functions dependent on $a$ and the endpoints $x_1,x_2$ of $A$. By the quasiparticle picture of local excitation \cite{He:2014mwa} the R\'enyi entropy for the superposition state $|\xi(k)\rangle$ takes the form
\be
\Delta S^{(n)}(\rho_A(k))=\frac{1}{1-n}\log\Big[\frac{|\alpha_k |^{2n}+|\beta_k|^{2n}}{(|\alpha_k |^2+|\beta_k|^2)^n}\Big].
\ee
By using the sum rule, we obtain the pseudo-R\'enyi entropy under the same condition
\bea
&&\phantom{=}\ep^{(1-n)\Delta S^{(n)}(\transA)}\nn \\
&&=\frac{1}{2n+1}\sum_{k=0}^{2n}\ep^{-\frac{2\pi \ii}{2n+1}k n} \frac{\ep^{(1-n)\Delta S^{(n)}(\rho_A(k))}}{[\mathcal{N}(\ep^{-\frac{2\pi \ii}{2n+1}k })]^{2n}\langle \phi| \psi\rangle^n}\nn \\
&&=\frac{1}{(2n+1)(1+\ep^{\ii\theta})^n}\sum_{k=0}^{2n}\ep^{-\frac{2\pi \ii}{2n+1}k n} (|\alpha_k |^{2n}+|\beta_k|^{2n})\nn\\
&&=\frac{1+\ep^{\ii n\theta}}{(1+\ep^{\ii\theta})^n}.
\eea
where in the second step we use the relation
\be
\mathcal{N}(\ep^{-\frac{2\pi \ii}{2n+1}k })^{2n}\langle \phi| \psi\rangle^n=(|a_k|^2+|\beta_k|^2)^{-n}(1+\ep^{\ii\theta})^n.\nn
\ee
We can predict that the pseudo-R\'enyi entropy  is given by
\be
\Delta S^{(n)}(\transA)=\frac{1}{1-n}\log \frac{1+\ep^{\ii n\theta}}{(1+\ep^{\ii\theta})^n}.
\ee
Remarkably, this prediction is consistent with the results obtained directly from calculations using the replica method. Furthermore, the aforementioned outcomes also align with the quasiparticle picture's explanation of the time evolution of the transition matrix.

Note that in \cite{Guo:2022sfl} the authors consider the time evolution of a superposition state and find the pseudo-R\'enyi entropy is associated with R\'enyi entropy in the later time limit. One could show the results in \cite{Guo:2022sfl} can be derived by using the sum rule (\ref{sumrulegeneral}) and quasiparticle picture \cite{He:2014mwa,Guo:2018lqq}.
~\\

\textit{Quasiparticle excited states in spin chain}\  Consider two distinct single-quasiparticle excited states denoted as $|k_1\rag$ and $|k_2\rag$ in spin chain. These states are orthogonal and pertain to a chain comprising $L$ free bosons or fermions. The system is divided into subsystem $A=[1,\ell]$ and its complementary $B=[\ell+1,L]$. The two states are defined by acting the mode operator $b_k=b_{A,k}+b_{B,k}$ on ground state
\be
|k\rag = b_{A,k}^\dag |G\rag+b_{B,k}^\dag |G\rag,
\ee
where $b_{A,k}$, $b_{B,k}$ are operators located in $A$ and $B$. It is shown in \cite{Castro-Alvaredo:2018dja,Zhang:2020vtc} the R\'enyi entropy has a universal picture, which can be explained as the probability of the quasiparticle located in $A$ is given by $\lag G| b_{A,k}b_{A,k}^\dagger|G\rangle=x:=\frac{\ell}{L}$. For the superposition state
\bea
&&|\xi(\ep^{\ii\phi})\rag = \f{1}{\sr2} ( |k_1\rag + \ep^{\ii\phi} |k_2\rag )\nn \\
&&=\f{1}{\sr2}\left[(b_{A,k_1}^\dagger+e^{i\phi}b_{A,k_2}^\dagger )|G\rangle+(b_{B,k_1}^\dagger+e^{i\phi}b_{B,k_2}^\dagger )|G\rangle\right].\nn
\eea
By similar argument the probability of the quasiparticle is located in $A$ is given by
\bea
&&p_{A,\xi}:=\frac{1}{2}\lag G| (b_{A,k_1}+e^{-i\phi}b_{A,k_2} )(b_{A,k_1}^\dagger+e^{i\phi}b_{A,k_2}^\dagger )|G\rag\nn \\
&&\phantom{p_{A,\xi}}=x+\frac{1}{2}\left(\a_{21}\ep^{-\ii\phi} + \a_{12}\ep^{\ii\phi}\right),
\eea
where
\be
\a_{12} = \ep^{-\f{\pi\ii k_{12} (\ell+1)}{L}} \f{\sin\f{\pi k_{12}\ell}{L}}{L\sin\f{\pi k_{12}}{L}}, ~~ k_{12}=k_1-k_2.
\ee
By this picture, the R\'enyi entropy takes the form
\bea \label{trArAxiepiithetanMT}
 \tr_A \r_{A,\xi(\ep^{\ii\phi})}^n = p_{A,\xi}^n+(1-p_{A,\xi})^n.
\eea
Using the sum rule (\ref{sumrulegeneral}) we can derive
\be \label{trArAk1k2nMT}
\tr_A (\cT_A^{k_2|k_1})^n = [1+(-)^n]\a_{12}^n.
\ee
This is consistent with the direct calculation.

\section{Tentative sum rule for pseudoentropy} \label{sectionTen}

For two nonorthogonal states $|\phi\rag$ and $|\psi\rag$, the pseudoentropy can be obtained by using
\be
\lim_{n\to 1} S^{(n)}(\transA) = - \partial_n \tr_A(\transA)^n |_{n \to 1}.
\ee
We want to expand the $S^{(n)}(\transA)$ and $S^{(n)}(\rho_A(k))$ in the sum rule near $n=1$.

The sum rule can be written as
\be
\tr_A (\transA)^n =\frac{1}{(2n+1)\langle \phi |\psi\rangle^n}\sum_{k=0}^{2n}\ep^{-\frac{2\pi \ii}{2n+1}k n} \frac{\tr_A\rho_A(k)^n}{\mathcal{N}(\ep^{\frac{2\pi \ii}{2n+1}k})^{2n}}.\nn
\ee
We expand $\tr_A (\transA)^n$ near $n\sim 1$ as
\be
\tr_A (\transA)^n\simeq 1 + (n-1) \partial_n \tr_A (\transA)^n|_{n=1}+O(n-1)^2.
\ee
Similarly, for $\tr_A\rho_A(k)^n$ we have
\be
\tr_A\rho_A(k)^n=1+(n-1)\partial_n \tr_A\rho_A(k)^n|_{n=1}+O(n-1)^2.
\ee
Using
\be
\frac{1}{(2n+1)\langle \phi |\psi\rangle^n}\sum_{k=0}^{2n}\ep^{-\frac{2\pi \ii}{2n+1}k n} \frac{1}{\mathcal{N}(\ep^{\frac{2\pi \ii}{2n+1}k})^{2n}}=1,
\ee
we have
\bea
&&\phantom{=}-\partial_n \tr_A \rho_A(k)^n|_{n=1} \\
&&= \sum_{k=0}^{2n} \frac{\ep^{-\frac{2\pi \ii}{2n+1}k n}}{(2n+1)\langle \phi |\psi\rangle^n} \frac{-\partial_n \tr_A\rho_A(k)^n|_{n=1}}{\mathcal{N}(\ep^{\frac{2\pi \ii}{2n+1}k})^{2n}}.\nn
\eea
Taking the limit $n\to 1$ on both side of the above equation, we get
\be
S(\transA)=\lim_{n\to 1}\sum_{k=0}^{2n}\frac{\ep^{-\frac{2\pi \ii}{2n+1}k n}}{3\langle \phi |\psi\rangle} \frac{S(\rho_A(k))|_{n=1}}{\mathcal{N}(\ep^{\frac{2\pi \ii}{2n+1}k})^{2n}}.
\ee
The reduced density matrix $\rho_A(k)$ depends on $n$. Generally, we have
\bea
&& \partial_n \tr_A \rho_A(k)^n=\partial_n \sum_i [\lambda_i(n)]^n \\
&& \hspace{6mm}
    = \sum_i \{ [\lambda_i(n)]^n \log \lambda_i(n)+n\lambda_i^{n-1}\partial_n \lambda_i(n) \},\nn
\eea
where $\lambda_i(n)$ denote the eigenvalues of $\rho_A(k)$.
Taking $n=1$ and using the fact that $\sum_i \lambda_i(n)=1$, we get the second term $\sum_i\lambda_i^{n-1}\partial_n \lambda_i(n)|_{n=1}=0$.

If one could take the limit first, we would have
\bea\label{SMsumpseudo}
S(\transA) = \sum_{k=0}^{2} \frac{\ep^{-\frac{2\pi \ii}{3}k }}{3\langle \phi |\psi\rangle} \frac{S(\rho_A(k))|_{n=1}}{[\mathcal{N}(\ep^{\frac{2\pi \ii}{3}k})]^{2}}.
\eea
But the sum rule (\ref{SMsumpseudo}) seems to be not true for the general case.

We consider the perturbation states. The pseudoentropy in this example can be evaluated by $S(\transA)=-\partial_n \tr_A (\transA)^n|_{n=1}$ 
and obtain
\bea
S(\transA)=S(\rho_A^\psi)+\epsilon^* \langle \psi' | [ K^\psi_A-S(\rho_A^\psi) ] |\psi\rangle,
\eea
which is the first law like relation for the pseudoentropy.
The entanglement entropy of the superposition state is
\bea
&& S(\rho_A(k))= S(\rho_A^\psi)
              + \frac{\epsilon^*}{1+\ep^{-\frac{2\pi \ii}{3}}}\langle \psi' | [ K^\psi_A-S(\rho_A^\psi) ]|\psi\rangle\nn \\
&&\phantom{S(\rho_A(k))=}
              + \frac{\epsilon}{1+\ep^{\frac{2\pi \ii}{3}}}\langle \psi | [K^\psi_A-S(\rho_A^\psi)] |\psi'\rangle.
\eea
Taking the above expression into RHS of (\ref{SMsumpseudo})  we have
\blea
&& \phantom{=}
\frac{1}{3}\sum_{k=0}^{2} \ep^{-\frac{2\pi \ii}{3}}|1+\ep^{-\frac{2\pi \ii}{3} k}|^2\Big[ S(\rho_A^\psi )+\frac{\epsilon}{1+\ep^{\frac{2 i \pi  k}{3}}} \langle \psi | K^\psi_A|\psi'\rangle+\frac{\epsilon^*}{1+\ep^{-\frac{2 i \pi  k}{3}}} \langle \psi' | [ K^\psi_A-S(\rho_A^\psi) ]|\psi\rangle+\frac{ \epsilon^*\langle \psi'| \psi\rangle S(\rho_A^\psi)}{1+\ep^{\frac{2 i \pi  k}{3}}}\Big]\nn \\
&&=S(\rho_A^\psi)+\epsilon^* \langle \psi' |[ K^\psi_A-S(\rho_A^\psi)]|\psi\rangle.
\elea
Therefore, the sum rule (\ref{SMsumpseudo}) is available for the perturbation states to the leading order of $\epsilon$. However, we do not expect it is still true to the next order.

\begin{figure}[htpb]
  \centering
  \includegraphics[scale=0.99]{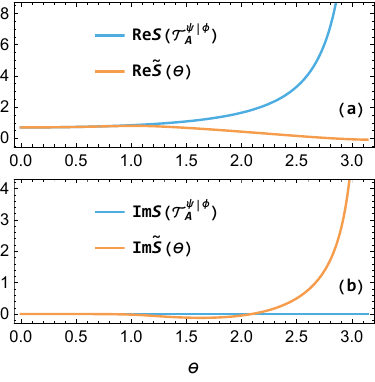}
  \caption{The real and imaginary parts of the function $\td{\mathcal{S}}(\theta)$ and pseudoentropy $S(\transA)$ as a function of $\theta$. In the region $\theta \ll 1$, $S(\transA) \approx \td\cS(\theta)$, which is consistent with the result that the pseudoentropy sum rule (\ref{SMsumpseudo}) is available in the leading order of the perturbation. }\label{SMF1}
\end{figure}

We consider the two qubits system as an example to show the sum rule for pseudoentropy (\ref{SMsumpseudo}) is not correct beyond the leading order of perturbation. For simplicity, we consider the state
\bea
&&|\phi\rangle= \frac{1}{\sqrt{2}}(|00\rangle +|11\rangle),\nn \\
&&|\psi\rangle =\frac{1}{\sqrt{2}}(|00\rangle +\ep^{\ii \theta}|11\rangle),
\eea
where $\theta\in [0,2\pi)$. The pseudoentropy is given by
\be\label{SMthetapseudo}
S(\transA)=\frac{1}{2} \theta  \tan \frac{\theta }{2}+\log \Big(2 \cos \frac{\theta }{2}\Big).
\ee
For the superposition state $|\xi(k)\rangle$ the entanglement entropy is
\bea
S(\rho_A(k))=-\lambda_1\log \lambda_1-\lambda_2\log \lambda_2,
\eea
where
\bea
&& \l_1 = 2 [ \mathcal{N}(\ep^{\frac{2\pi \ii}{2n+1}k})]^2 \cos^2\Big(\frac{\pi  k}{3}\Big), \nn\\
&& \l_2 = 2 [ \mathcal{N}(\ep^{\frac{2\pi \ii}{2n+1}k})]^2 \cos^2\Big(\frac{\theta }{2}+\frac{\pi  k}{3}\Big), \\
&& [ \mathcal{N}(\ep^{\frac{2\pi \ii}{2n+1}k})]^2 =
    \frac{1}{2 [ \cos^2(\frac{\pi  k}{3}) + \cos^2(\frac{\theta }{2}+\frac{\pi  k}{3}) ]}. \nn
\eea
We define the function
\bea
\td{\mathcal{S}}(\theta)=\frac{1}{3\langle \phi |\psi\rangle}\sum_{k=0}^{2}\ep^{-\frac{2\pi \ii}{3}k } \frac{S(\rho_A(k))|_{n=1}}{[\mathcal{N}(\ep^{\frac{2\pi \ii}{3}k})]^{2}}.
\eea
One could check $\tilde{S}(\theta)$ is different from the pseudoentropy (\ref{SMthetapseudo}) for arbitrary $\theta$. In fact for small $\theta\ll 1$ we have the expansion
\bea
&&S(\transA)=\log (2)+\frac{\theta ^2}{8}+\frac{\theta ^4}{64}+O\left(\theta ^6\right), \\
&& \td{\mathcal{S}}(\theta)=\log (2)+\frac{\theta ^2}{8}+\frac{\theta ^4}{192}-\frac{\ii \theta ^5}{64}+O\left(\theta ^6\right). \nn
\eea
We show $\tilde{S}(\theta)$ and $S(\transA)$ as function of $\theta$ in Fig.\ref{SMF1}.

In summary, the sum rule formula for pseudoentropy (\ref{SMsumpseudo}) is true to the leading order of the perturbation parameter $\epsilon$. However, it cannot be used beyond the leading order of $\epsilon$.
One possible explanation for the above conclusion is that the pseudoentropy and entanglement entropy both satisfy first-law like relation at the leading order of $\epsilon$, which are associated with the expectation value of the modular Hamiltonian.
Therefore, we conclude that (\ref{SMsumpseudo}) can only be used if the distance between the two states $\ketphi$ and $\ketpsi$ is small.
It would be interesting to find the sum rule for pseudoentropy for general cases.

\section{Alternative forms of the sum rule} \label{sectionAlt}

For the one-point function $\langle \phi| \mA |\psi\rangle$, we can also construct the sum rule by the following way. Let us write down the formula
\bea
\braphi \mA \ketpsi= \sum_{c\in \{1,-1,\ii,-\ii\}} a_{(c)}\mA_{(c)},\nn
\eea
where $a_{(c)}=2c+\braphi \psi\rangle+\brapsi \phi\rangle$ and
\bea
&&\mA_{(c)}:= \langle \xi_c| \mA |\xi_c\rangle\nn \\
&&|\xi_c\rangle=\frac{1}{\sqrt{1+c\braphi \psi\rangle +c^* \brapsi \phi\rangle+|c|^2}}(|\phi\rangle+ c |\psi\rangle).\nn
\eea
Similar formula has been used in \cite{Lashkari:2015dia} to calculate the off-diagonal elements of the modular Hamiltonian.

Now consider the other form of sum rule for pseudo-R\'enyi entropy. Consider $\tr_A(\transA)^2$ as an example. For simplicity, let $\langle \phi |\psi\rangle=0$. One could choose $\tilde{\mathcal{S}}=\left\{1,\ii,-1,-\ii,\ep^{\ii \theta }\right\}$, where $\theta$ is some constant. Define the superposition state $|\xi_c\rangle=\frac{1}{\sqrt{1+|c|^2}}(|\phi\rangle+ c| \psi\rangle)$. The sum rule can be expressed as
\bea\label{SMaltsum}
\tr_A(\transA)^2=\sum_{c\in \tilde{\mathcal{S}}} a_{(c)} \tr_A (\rho_{A,c})^2,
\eea
where $\rho_{A,c}:=\tr_{\bar A}|\xi_c\rangle \langle \xi_c| $, $a_c$ are given by
\bea
&& a_{(1)}=\frac{1}{1-\ep^{\ii \theta }},\
   a_{(\ii)}=-\frac{1}{1+\ii \ep^{i \theta }},\
   a_{(-1)}=\frac{1}{1+\ep^{i \theta }},\nn \\
&& a_{(-\ii)}=-\frac{1}{1-\ii \ep^{\ii \theta }},\
   a_{(\ep^{\ii\theta})}=-\frac{4 \ep^{2 \ii \theta }}{1-\ep^{4 \ii \theta }}.
\eea

Another form is by  taking the set $\tilde{\mathcal{S}}'=\left\{1,\ep^{\ii \theta },\ep^{2 \ii \theta },\ep^{3 \ii \theta },\ep^{4 \ii \theta }\right\}$. The superposition state is $|\xi_c\rangle= \frac{1}{\sqrt{1+|c|^2}}(|\phi\rangle+c |\psi\rangle)$.
We have the relation
\bea
\tr_A(\transA)^2 =\sum_{c\in \tilde{\mathcal{S}}'} a_{(c)} \tr_A (\rho_{A,c})^2,
\eea
where
\bea
&& a_{(1)} = \frac{4}{(1-\ep^{\ii \theta })^4 (1+\ep^{\ii \theta })^2 (1+\ep^{\ii \theta }+2 \ep^{2 \ii \theta }+\ep^{3 \ii \theta }+\ep^{4 \ii \theta })},\nn \\
&& a_{(\ep^{\ii\theta})}=-\frac{4 \ep^{-\ii \theta }}{(1-\ep^{\ii \theta })^4 (1+\ep^{\ii \theta }) (1+\ep^{\ii \theta }+\ep^{2 \ii \theta })},\nn \\
&& a_{(\ep^{2\ii\theta})}= \frac{4 \ep^{-\ii \theta }}{(1-\ep^{\ii \theta })^4 (1+\ep^{\ii \theta })^2},\nn \\
&& a_{(\ep^{3\ii\theta})}=-\frac{4}{(1-\ep^{\ii \theta })^4 (1+\ep^{\ii \theta }) (1+\ep^{\ii \theta }+\ep^{2 \ii \theta })},\nn \\
&& a_{(\ep^{4\ii\theta})}=\frac{4 \ep^{2 \ii \theta }}{(1-\ep^{\ii \theta })^4 (1+\ep^{\ii \theta })^2 (1+\ep^{\ii \theta }+2 \ep^{2\ii \theta }+\ep^{3 \ii \theta }+\ep^{4 \ii \theta })}.\nn
\eea
One could construct other forms of sum rule by choosing different set $\mathcal{S}$.
But these solutions are only for some special cases.

\section{Generalized R\'enyi entropy as derivatives of R\'enyi entropy}\label{sectionGen}

In \cite{Murciano:2021dga} the authors introduce the so-called the generalized R\'enyi entropy, akin to the pseudo-R\'enyi entropy. In this section, we aim to explore the potential to derive the generalized R\'enyi entropy through derivatives of the R\'enyi entropy.
In a $d$-dimensional Hilbert space, we define the mixed state reduced density matrix
\be
\r_A(\{c_i\}) = \sum_{i=1}^d c_i \r_{A,i},
\ee
with the eigenstate reduced density matrices
\be
\r_{A,i} = \tr_{\bar A} |i\rag\lag i|.
\ee
Here the reduced density matrices $\r_A(\{c_i\})$ and all $\r_{A,i}$ are not necessarily well normalized.
For $\r_A(\{c_i\})$ to be a well-defined (unnormalized) reduced density matrix, we require $c_i\geq0$ and $\sum_{i=1}^d c_i>0$.
From
\bea
&& [\r_A(\{c_i\})]^n = \sum_{\{0\leq r_i \leq n\}}^{\sum_{i=1}^d r_i=n}
                    \Big\{ \Big( \prod_{i=1}^d c_i^{r_i} \Big)\\
&& \phantom{[\r_A(\{c_i\})]^n =} \times
                           \Big[ \Big( \prod_{i=1}^d \r_{A,i}^{r_i} \Big) + \cdots \Big]_{\f{n!}{\prod_{i=1}^d (r_i!)}} \Big\},\nn
\eea
for a set $\{r_i\}$ with $0\leq r_i \leq n$ and $\sum_{i=1}^d r_i=n$, we get
\be \label{SMProductOfRDMs}
\Big[ \Big( \prod_{i=1}^d \r_{A,i}^{r_i} \Big) + \cdots \Big]_{\f{n!}{\prod_{i=1}^d (r_i!)}} =
\Big[ \prod_{i=1}^d \f{\p_{c_i}^{r_i}}{r_i!} \Big] [\r_A(\{c_i\})]^n.
\ee
In the above two equations we have used `$\cdots$' to denote the permutation terms, and subscript $\f{n!}{\prod_{i=1}^d (r_i!)}$ denotes the total number of terms.
In this way, we the sum of proper products of the reduced density matrices as the derivatives of the power of one reduced density matrix.

More generally, we define the mixed state reduced density matrix
\be
\r_A(\{c_{i,i'}\}) = \sum_{i,i'} c_{i,i'} \r_{A,i,i'},
\ee
with the reduced transition matrices
\be
\r_{A,i,i'} = \tr_{\bar A} |i\rag\lag i'|.
\ee
For $\r_A(\{c_{i,i'}\})$ to be a well-defined (unnormalized) reduced density matrix, we require that it is Hermitian and positive definite.
Especially, we require $c_{i,i'}^*=c_{i',i}$, $c_{i,i}\geq0$ and $\sum_i c_{i,i}>0$.
From
\bea
&& [\r_A(\{c_{i,i'}\})]^n = \sum_{\{0\leq r_{i,i'}\leq n\}}^{\sum_{i,i'}r_{i,i'}=n}
                         \Big\{ \Big(\prod_{i,i'}c_{i,i'}^{r_{i,i'}}\Big) \\
&& \phantom{[\r_A(\{c_{i,i'}\})]^n =} \times
                         \Big[ \Big( \prod_{i,i'} \r_{A,i,i'}^{r_{i,i'}} \Big) + \cdots \Big]_{\f{n!}{\prod_{i,i'} (r_{i,i'}!)}} \Big\}, \nn
\eea
we get the sum of proper products of the reduced transition matrices written as the derivatives of the power of one reduced density matrix
\be \label{SMProductOfGRDMs}
\Big[ \Big( \prod_{i,i'} \r_{A,i,i'}^{r_{i,i'}} \Big) + \cdots \Big]_{\f{n!}{\prod_{i,i'} (r_{i,i'}!)}} =
\prod_{i,i'} \f{\p_{c_{i,i'}}^{r_{i,i'}}}{r_{i,i'}!} [\r_A(\{c_{i,i'}\})]^n.
\ee

\section{Discussion} \label{sectionDis}

In this paper, we have introduced a novel operator sum rule that involves the reduced transition matrix $\trans$ and the density matrix of the superposition state $|\xi(c)\rangle$. Utilizing this sum rule, we find the off-diagonal matrix elements can be associated with the diagonal matrix elements (\ref{correlator_sumrule_nonorth})(\ref{correlator_sumrule_orth}). We also establish a connection between pseudo-R\'enyi entropy and R\'enyi entropy (\ref{sumrulegeneral}), both of which have significant physical implications.

Currently, the physical interpretation of the non-Hermitian transition matrix and pseudo-R\'enyi entropy remains unclear, despite some intriguing findings in recent research. Our sum rule can be viewed as a bridge linking these new concepts to established ones.
We demonstrate the significance of the sum rules and their potential applications across various physics domains, including understanding the gravity dual of the non-Hermitian transition matrix and the quasi-particle interpretation of pseudo-R\'enyi entropy. It is worthwhile to further explore the physics applications of the sum rule.

Constructing a sum rule for pseudoentropy may be particularly intriguing, as the pseudoentropy is also expected to satisfy the HRT formula within the framework of AdS/CFT. The corresponding sum rule could potentially offer insightful geometric interpretations. The sum rule for the pseudo-R\'enyi entropy presented in this paper does not exhibit a smooth limit to the pseudoentropy. Thus, a sum rule for the pseudoentropy is still lacking. However, in principle, there should be various forms of the sum rule achievable by selecting the set $\mathcal{S}$. It is possible to construct one that could have a smooth limit for the pseudoentropy.

Moreover, we anticipate that our findings can be extended to the transition matrix for mixed states \cite{Nakata:2020luh,Guo:2022jzs} and the generalized R\'enyi entropy \cite{Murciano:2021dga}.
Notably, a recent work \cite{Parzygnat:2023avh} introduced a new quantity called SVD entanglement entropy, utilizing the transition matrix, which exhibits distinct properties compared to pseudoentropy. It would be fascinating to investigate whether the SVD entanglement entropy also possesses a sum rule similar to the pseudo-R\'enyi entropy.

~\\
{\bf Acknowledgements}
We would like to thank Song He and  Yu-Xuan Zhang for helpful discussions.
WZG is supposed by the National Natural Science Foundation of China under Grant No. 12005070 and the Fundamental Research Funds for the Central Universities under Grants No. 2020kfyXJJS041.
JZ acknowledges the support from the National Natural Science Foundation of China (NSFC) through Grant No. 12205217.



\providecommand{\href}[2]{#2}\begingroup\raggedright\endgroup

\end{document}